\documentclass[10 pt, journal]{IEEEtran} 
\usepackage{hyperref}

\usepackage{multirow}

\usepackage{amsmath}
\usepackage{amsthm}
\usepackage{amssymb}

\usepackage{algorithmicx}
\usepackage{algorithm}
\usepackage{algpseudocode}

\usepackage{graphicx}
\usepackage{tikz}

\usepackage{array}

\newtheorem{definition}{Definition}
\newtheorem{remark}{Remark}
\begin{document}
\title{Validation methodology on real data of reversible Kalman Filter for state estimation with Manifold}

%\author{
%\IEEEauthorblockN{Svyatoslav Covanov}
%\IEEEauthorblockA{
%IRL2958 GT-CNRS\\
%GeorgiaTech Europe - Metz, France\\
%svyatoslav.covanov@polytechnique.org}\\
%\bigskip
%\IEEEauthorblockN{C\'edric Pradalier}
%\IEEEauthorblockA{
%IRL2958 GT-CNRS\\
%GeorgiaTech Europe - Metz, France\\
%cedric.pradalier@georgiatech-metz.fr}
%}

%\author{
%\IEEEauthorblockN{Svyatoslav Covanov\IEEEauthorrefmark{1} \quad
%Cédric Pradalier\IEEEauthorrefmark{2}}\\
%\IEEEauthorblockA{\IEEEauthorrefmark{1}IRL2958 GT-CNRS, GeorgiaTech Europe, Metz, France\\
%Email: svyatoslav.covanov@polytechnique.org}\\
%\IEEEauthorblockA{\IEEEauthorrefmark{2}IRL2958 GT-CNRS, GeorgiaTech Europe, Metz, France\\
%Email: cedric.pradalier@georgiatech-metz.fr}
%}
\author{
\begin{minipage}[t]{0.48\textwidth}
\centering
Svyatoslav Covanov\\
IRL2958 GT-CNRS\\
GeorgiaTech Europe - Metz, France\\
svyatoslav.covanov@polytechnique.org
\end{minipage}%
\hfill
\begin{minipage}[t]{0.48\textwidth}
\centering
Cédric Pradalier\\
IRL2958 GT-CNRS\\
GeorgiaTech Europe - Metz, France\\
cedric.pradalier@georgiatech-metz.fr
\end{minipage}
}

\markboth{Journal of \LaTeX\ Class Files,~Vol.~14, No.~8, August~2015}%
{Shell \MakeLowercase{\textit{et al.}}: Bare Demo of IEEEtran.cls for IEEE Journals}



%\algrenewcommand\Return{\State \textbf{return }}
\maketitle

%\newcommand\submittedtext{%
%  \footnotesize This work has been submitted to the IEEE for possible publication. Copyright may be transferred without notice, after which this version may no longer be accessible.}
%
%\newcommand\submittednotice{%
%\begin{tikzpicture}[remember picture,overlay]
%\node[anchor=south,yshift=10pt] at (current page.south) {\fbox{\parbox{\dimexpr0.65\textwidth-\fboxsep-\fboxrule\relax}{\submittedtext}}};
%\end{tikzpicture}%
%}

\begin{abstract}
This work extends a previous study that introduced an algorithm for state estimation on manifolds within the framework of the Kalman filter. Its objective is to address the limitations of the earlier approach. The reversible Kalman filter was designed to provide a methodology for evaluating the accuracy of existing Kalman filter variants with arbitrary precision on synthetic data. It has favorable numerical properties on synthetic data, achieving arbitrary precision without relying on the small-velocity assumption and depending only on sensor noise. However, its application to real data encountered difficulties related to measurement noise, which was mitigated using a heuristic. In particular, the heuristic involved an event detection step switching between reversible Kalman filter and classical Kalman variant at chosen moments. In the present work, we propose a study of this detection step and propose a methodology to prove at which moment the reversible Kalman approach improves on classical multiplicative variant.
\end{abstract}


% \fi
%\IEEEpeerreviewmaketitle

%\submittednotice
\section{Introduction}

{T}{his} manuscript substantially extends the recent work by Covanov and Pradalier in~\cite{covanov2025reversiblekalmanfilterstate} by reusing the concept of a reversible Kalman filter and proposing a method to validate it on real data.
This study is motivated by applications in the routine inspection of large metallic structures, such as ship hulls, in underwater environments. Its aim is to improve the localization of a differential-drive robot equipped with acoustic sensors, used for the inspection of ship surfaces. The work focuses on enhancing the fusion filter involved in computing the robot’s orientation and position by incorporating knowledge of the normal vector of the surface on which the robot moves. The approach was first developed theoretically and applied to synthetic data, then generalized to real-world scenarios.

A general discussion of state estimation in robotics can be found in~\cite{10.5555/3165227}.
Research on state estimation on manifolds has only recently gained attention. One of the first efforts in this direction was presented in~\cite{starbuck:hal-03445976}, where a method derived from the Invariant Extended Kalman Filter (IEKF), as described in~\cite{annurev:/content/journals/10.1146/annurev-control-060117-105010}, was applied to localize a magnetic crawler on a ship hull. This work proposed a rigorous mathematical framework that relies on wheel encoders for the prediction step and ultra-wideband measurements for the correction step. This approach followed the idea of implementing consistent variants of Kalman filters that preserve the manifold structure, as described in~\cite{phdthesisReady,Barrau2015AnEA}.

Additional efforts have recently emerged based on the frameworks developed in~\cite{Mahony2020,Mahony2022,FORNASIER2025112495}. These works exploit system symmetries, giving rise to modern equivariant frameworks. This line of research builds on decades of studies on symmetry-preserving estimation, notably through the development of the invariant Kalman filter~\cite{Bonnabel2009}. Very recent studies have sought to integrate manifold representations into these approaches, as in~\cite{ge2023noteextendedkalmanfilter,ge2025geometryextendedkalmanfilters}.

In our context, by design, the sensors embedded in an IMU, such as gyroscopes, accelerometers or magnetomers, require a three-dimensional state representation, whereas wheel encoders encode a two-dimensional state vector (which, although embedded in 3D, lies on a 2D differentiable manifold). This creates a challenge of consistently projecting IMU measurements onto a 2D manifold while preserving the covariance structure of the Kalman filter. For instance, for motion constrained to a plane orthogonal to the IMU’s $z$-axis, small rotations around the $x$ and $y$ axes may occur. Such variations will not be captured by the fusion filter if the gyroscope projection is limited to the $z$-axis. Moreover, accelerometer and magnetometer measurements must be appropriately rotated so that the Kalman update remains consistent with the 2D state formulation. Thus, the question arises of how to incorporate surface information in the most natural and consistent way.

This problem raises two main challenges. First, accelerometers measure both gravity and external accelerations, making the separation of these components nontrivial. Several recent works address this issue through signal-processing techniques that filter specific components of the accelerometer signal~\cite{Lee2012EstimationOA,Widodo2016}. Second, magnetometer readings are often disturbed, particularly near ship hulls. A common mitigation strategy is to incorporate pressure sensors as complementary sources of information.

The approach introduced in~\cite{covanov2025reversiblekalmanfilterstate}, which forms the basis of this work, is complementary to these efforts and focuses on geometric constraints. Instead of embedding Kalman filtering within a general symmetry-based framework, it analyzes the symbolic structure of constraints arising from overdetermined sensor configurations. This leads to the following theoretical result: in the absence of magnetic disturbances and under surface-constrained motion, external accelerations can be eliminated with arbitrary precision on synthetic data. Such constraints compensate for traditional IMU shortcomings by adding information from surface or pressure sensors. As a consequence, the method improves the testability of the Kalman filter and ensures that numerical precision on synthetic data does not depend on data variability. However, its generalization to real-world data was not fully explored.

This observation motivates a deeper study of the filter’s behavior on real datasets. The original implementation of the reversible Kalman filter relied on a heuristic designed to indicate when geometric constraint resolution should be applied. In this work, we evaluate this approach on several datasets and use a metric capable of identifying time periods where the reversible Kalman filter outperforms the classical MEKF.
Identifying an appropriate metric is essential, as practical cases in which the reversible approach is superior are difficult to demonstrate: the MEKF typically performs well on real data for orientation estimation, whereas the reversible approach exhibits favorable asymptotic behavior in theory. We therefore bridge this gap and enable the practical application of the theoretical reversible Kalman filter to real-world scenarios.

This opens the door to applications such as trajectory estimation for vehicles on flat or nearly flat surfaces (e.g., ground robots or human motion analysis) and underwater navigation, where the robot’s orientation remains approximately normal to the surface, allowing pressure sensors to substitute for magnetometers.

In the present work, we first detail the mathematical properties of the reversible Kalman filter and discuss its theoretical characteristics. We then examine the conditions under which it can be applied to real datasets. We analyze the behavior of the original heuristic and propose a way to mitigate its limitations by identifying where it provides improvements over MEKF. This comparison enables an automatic, human-inspired decision process in the correction phase.

The paper is organized as follows, using $9$-axis IMU sensors as the main example. Section~\ref{sec:backtheory} introduces the notation and provides a theoretical overview of quaternion properties. Section~\ref{sec:mekf} presents the Multiplicative Extended Kalman Filter (MEKF) used as a baseline. Section~\ref{sec:revmekf} describes the reversible Kalman filter and evaluates it on synthetic data, introducing the metric used throughout the rest of the paper. Section~\ref{sec:evalmekf} reviews the heuristic from~\cite{covanov2025reversiblekalmanfilterstate}. Section~\ref{sec:datasets} compares this heuristic and MEKF on several real datasets. Finally, Section~\ref{sec:discussions} discusses the main applications and limitations of the proposed approach.

The code for this project is available in the following repository:
\href{https://github.com/knov836/revmekf}{https://github.com/knov836/revmekf}.

\section{Background Theory}
\label{sec:backtheory}

In this section, we introduce the notations and mathematical background necessary for formulating the algorithms based on the multiplicative Kalman filter, described in the following sections.  
In our approach, maintaining arbitrary numerical precision is the main criterion. Therefore, a rigorous modeling of the sensors and the corresponding rotations is required.

\subsection{Quaternions, Rotations, and Axis–Angle Representation}

A detailed exposition of quaternion kinematics can be found in~\cite{DBLP:journals/corr/abs-1711-02508}.  
In our context, we consider the orientation of an IMU, which can be interpreted as a rotation of the $3$D space in the global Earth frame. Thus, it corresponds to elements of $SO(3)$. There exists a well-known correspondence, detailed in~\cite{Diebel2006RepresentingA}, between unit quaternions and elements of $SO(3)$.

First, quaternions $q \in \mathbb{H}$ can be represented as $4$D vectors:
\[
q = [q_0, q_1, q_2, q_3]^T = 
\begin{pmatrix}
q_0 \\ 
\mathbf{q}_{1:3}
\end{pmatrix}.
\]

The quaternion multiplication can be defined through an associated multiplication matrix $Q(q)$:
\[
Q(q) =
\begin{pmatrix}
	q_0 & -q_1 & -q_2 & -q_3\\
	q_1 & q_0  & q_3  & -q_2\\
	q_2 & -q_3 & q_0  & q_1\\
	q_3 & q_2  & -q_1 & q_0
\end{pmatrix}.
\]
Thus, for any two quaternions $q_1$ and $q_2$, one has  
\[
Q(q_1)\cdot Q(q_2) = Q(q_1 \cdot q_2).
\]

For a rotation $R \in SO(3)$ and a vector $v \in \mathbb{R}^3$, we denote by $R \cdot v$ the action of $R$ on $v$.  
Let $(0|v)$ denote the quaternion associated with $v$ by appending a fourth coordinate equal to zero.  
For a unit quaternion $q$, one has:
\[
q \cdot (0|v) \cdot q^{-1} = Q(q)^T \cdot Q(q) \cdot (0|v)
= 
\begin{pmatrix}
1 & 0\\
0 & R(q)
\end{pmatrix} \cdot (0|v),
\]
where $R(q)$ is the rotation matrix corresponding to $q$.  
This relation establishes the correspondence between unit quaternions and rotations.  

For a vector $v$, we simplify the notation of the quaternion action by writing simply  
\[
q^{-1} \cdot v \cdot q,
\]
and we denote by $v \cdot q$ the quaternion product $(0|v) \cdot q$.

Given a $3$D vector $w = [w_x, w_y, w_z]^T$, one can associate to it the skew-symmetric matrix $[w]_\times$:
\[
[w]_\times =
\begin{pmatrix}
	0 & -w_z & w_y\\
	w_z & 0 & -w_x\\
	-w_y & w_x & 0
\end{pmatrix}.
\]
This matrix satisfies $[w]_\times \cdot v = w \times v$, i.e., the cross product between $w$ and $v$.  
This operation is useful for determining the rotation axis between two vectors.

Furthermore, unit quaternions form a Lie group~\cite{sol2021microlietheorystate}, a property that is particularly useful in our application, since we make use of quaternion derivatives. The natural way to handle these operations is through the Lie group structure.  
In particular, unit quaternions admit a logarithmic map, providing a correspondence between unit quaternions and elements of $\mathbb{R}^3$:  
\[
\log(q) = u = \alpha \mathbf{u},
\]
where $\mathbf{u}$ is a unit vector representing the rotation axis, and $\alpha$ is the rotation angle.

Conversely, the exponential map from $\mathbb{R}^3$ to the set of unit quaternions is given by  
\[
\exp(\alpha \mathbf{u}) = 
\begin{pmatrix}
\cos\left(\frac{\alpha}{2}\right)\\
\mathbf{u}\sin\left(\frac{\alpha}{2}\right)
\end{pmatrix}.
\]
This representation corresponds to the following rotation matrix:
\[
R(\exp(\alpha \mathbf{u})) = 
I_3 + \sin(\alpha)\,[\mathbf{u}]_\times + (1 - \cos(\alpha))\,[\mathbf{u}]_\times^2.
\]
\subsection{IMU Measurement Model}

Given our IMU-sensor, one needs a correct model of its behavior in order to achieve an arbitrary precision. In our application, we consider a standard IMU composed of a gyroscope, an accelerometer and a magnetometer.
The gyroscope provides angular velocity measurements. Thus, at sample $k$, its output is a $3D$ vector $\boldsymbol{\omega}_k = [\omega_{k,x},\omega_{k,y},\omega_{k,z}]$.

One needs a precise relation between $\boldsymbol{\omega}_k$ and $q_k$ and $q_{k+1}$, which will define the way we construct synthetic data.  In~\cite{DBLP:journals/corr/abs-1711-02508,DBLP:journals/corr/Zhao16c}, one has the following relations in the continuous version of the problem of integrating gyroscopic data:
\begin{equation}
 \dot{q}^\mathcal{G} = \frac{1}{2}\boldsymbol{\omega}^\mathcal{G}\cdot q^\mathcal{G}
\end{equation}
where
\begin{equation}
	\boldsymbol{\omega}_\mathcal{G}(t) = \lim_{\Delta t \rightarrow 0}\frac{\log(q^\mathcal{G}(t+\Delta t)\cdot (q^\mathcal{G}(t))^{-1})}{\Delta t}
\end{equation}
$\mathcal{G}$ denoting the global frame and $\mathcal{R}$ the relative frame. From that relation, one obtains:
\begin{equation}
\dot{R}^\mathcal{G}(t) = [\boldsymbol{\omega}^\mathcal{G}]_\times R^\mathcal{G}(t).
\end{equation}
One has to be careful that these relations are in global frame. If the angular perturbation is given in relative frame, then the relation is:
\begin{equation}
	\label{eq:cont_quat_diff}
 \dot{q}^\mathcal{G} = \frac{1}{2}q^\mathcal{G}\otimes \boldsymbol{\omega}^\mathcal{R}
\end{equation}
and
\begin{equation}
\dot{R}^\mathcal{G}(t) = R^\mathcal{G}(t) [\boldsymbol{\omega}^\mathcal{R}]_\times.
\end{equation}
Since we obtain gyroscopic data in relative frame, given as an axis and a rotation speed in $\text{rad}.\text{s}^{-1}$, we use the last formulas.

From previous formulas, one can deduce an expression in discrete form. We choose a model where between samples $k$ and $k+1$ the angular rate $\omega_k$ is constant, which is an approximation allowing us to do zero-th order integration of the previous formulas.
The continuous-time differential equation~\ref{eq:cont_quat_diff} can then be discretized into
\begin{equation}
q_{k+1}^{\mathcal{G}} = q_{k}^{\mathcal{G}} \cdot \exp\big(\boldsymbol{\omega}_k (t_{k+1}-t_k)\big).
\end{equation}
Thus, for a quaternion $q_k$ representing orientation at time step $k$, the gyroscope yields a 3D vector $\boldsymbol{\omega}_k$ such that  
\begin{equation}
q_{k}^{\mathcal{G}} = q_{k-1}^{\mathcal{G}}\cdot \exp({\boldsymbol{\omega}_k}\cdot dt)
\end{equation}
	and
\begin{equation}
q_{k}^{\mathcal{R}} = \exp(-{\boldsymbol{\omega_k}}\cdot dt)\cdot q_{k-1}^{\mathcal{R}}.
\end{equation}

For $k > 0$, the relation between successive quaternions $q_{k+1}^{\mathcal{G}}$ and $q_{k}^{\mathcal{G}}$ is given by  
\begin{equation}
\log\left(\left(q_{k}^{\mathcal{G}}\right)^{-1} \cdot q_{k+1}^{\mathcal{G}}\right) = \boldsymbol{\omega}_k \, \Delta t.
\end{equation}

Equivalently,  
\begin{equation}
q_{k+1}^{\mathcal{G}} = q_{0}^{\mathcal{G}} \cdot \prod_{i=0}^{k} \left( \left(q_{i}^{\mathcal{G}}\right)^{-1} \cdot q_{i+1}^{\mathcal{G}} \right).
\end{equation}

The global quaternion enables the transformation of sensor measurements from the relative frame to the global frame. For the accelerometer, this yields  
\begin{equation}
q_{k}^{\mathcal{G}} \cdot \mathbf{A}_k^{\mathcal{R}} \cdot \big(q_{k}^{\mathcal{G}}\big)^{-1} = \mathbf{A}_k^{\mathcal{G}}.
\end{equation}
In the absence of external accelerations, $\mathbf{A}_k^{\mathcal{G}}$ is a noisy estimate of the gravity vector.  

Similarly, for the magnetometer,  
\begin{equation}
q_{k}^{\mathcal{G}} \cdot \mathbf{M}_k^{\mathcal{R}} \cdot \big(q_{k}^{\mathcal{G}}\big)^{-1} = \mathbf{M}_k^{\mathcal{G}},
\end{equation}
where $\mathbf{M}_k^{\mathcal{G}}$ is a measurement of the Earth’s magnetic field.  

To summarize, the available IMU outputs are $\boldsymbol{\omega}^{\mathcal{R}}_k$, $\mathbf{A}_k^{\mathcal{R}}$, and $\mathbf{M}_k^{\mathcal{R}}$.

\section{Multiplicative Extended Kalman Filter (MEKF)}
\label{sec:mekf}

The Multiplicative Kalman Filter is a variant of the standard Kalman filter~\cite{10.1115/1.3662552}, in which operations are performed using the $4D$ structure of $\mathbb{H}$, without accounting for the fact that we are working with unit quaternions. The multiplicative Kalman variant was described in~\cite{Maley2013MultiplicativeQE} and preserves the Lie group structure of unit quaternions.

Although this approach is well known, several implementation options exist. In particular, it is necessary to specify the choice of residual used in the update step. Here, we describe the variant that has been implemented, which defines the structure of the inner state.

\subsection{Structure of the algorithm}
In our context, we assume that the Multiplicative Extended Kalman Filter (MEKF) takes as input a sequence of sensor measurements 
$\boldsymbol{\omega}_k^{\mathcal{R}}$, $\boldsymbol{A}_k^{\mathcal{R}}$, and $\boldsymbol{M}_k^{\mathcal{R}}$.

The general structure of the implemented MEKF variant is as follows: at each time step $k$, MEKF updates an inner state vector $\hat{x}_k$ in two steps: a prediction step and an update step.  
The inner state $\hat{x}_k$ is a 6-dimensional vector, where the first three components correspond to the logarithm of a unit quaternion (axis–angle representation), and the last three components correspond to the gyroscope bias. This variant allows correction of the gyroscope bias and helps study the effect of sensor noise on bias estimation.

The first three components of the inner state form a vector that allows computing a quaternion $q_{k}$ using:
\[
q_{k} = \exp(\hat{x}_{k}[1:3]),
\]
where the coefficient indices start at $1$ (not $0$).

At time step $k$:  
\begin{itemize}
	\item \textbf{Prediction step:} Compute $q_{k|{k-1}}$ using the gyroscope data $\boldsymbol{\omega}_k$, and update
\begin{align*}
	\hat{x}_k[1:3] &\leftarrow \log(q_{k|k-1}), \\
    \hat{x}_k[4:6] &\leftarrow \hat{x}_k[4:6].
\end{align*}

	\item \textbf{Update step:} MEKF corrects the inner state vector using the accelerometer and magnetometer measurements. From two independent vectors $(\mathbf{A}_k^\mathcal{R},\mathbf{M}_k^\mathcal{R})$ and their corresponding directions in the global frame $(\mathcal{A}^\mathcal{G},\mathcal{M}^\mathcal{G})$, there exists a unique proper rotation that maps the first pair to the second.

	In static conditions, the uniqueness of this rotation follows from the identity:
$q_{k}^{\mathcal{G}}\cdot \left(\mathbf{A}_k^{\mathcal{R}}\times \mathbf{M}_k^{\mathcal{R}}\right)\cdot \left(q_{k}^{\mathcal{G}}\right)^{-1} = \left(q_{k}^{\mathcal{G}}\cdot \mathbf{A}_k^{\mathcal{R}}\left(q_{k}^{\mathcal{G}}\right)^{-1}\right) \times \left(q_{k}^{\mathcal{G}}\cdot \mathbf{M}_k^{\mathcal{R}}\left(q_{k}^{\mathcal{G}}\right)^{-1}\right),$
because this relation fixes the third vector of a basis of $\mathbb{R}^3$ starting with the $2$ vectors $\mathcal{A}^\mathcal{R}$ and $\mathcal{M}^\mathcal{R}$. Another consequence is that, in non static conditions, this relation is an approximation of this unique rotation.
\end{itemize}

\begin{algorithm}
\caption{Kalman Filter: Prediction and Update}\label{alg:Kalman}
\begin{algorithmic}[1]

	\Require Previous state estimate $\hat{x}_{k-1}$, covariance $P_{k-1}$, gyroscope $\boldsymbol{\omega} = (p, q, r)$, accelerometer $\boldsymbol{A}$, magnetometer $\boldsymbol{M}$, process noise matrix $Q_k$, measurement noise matrix $U_k$, time step $dt$
	\Ensure Updated state estimate $\hat{x}_k$, covariance $P_{k}$

	\Statex $\hat{x}_{k|k-1},P_{k|k-1} \leftarrow$\textbf{Prediction}($\hat{x}_{k-1},P_{k-1},Q_k,\boldsymbol{\omega},dt$)
	\Statex $\hat{x}_{k},P_{k} \leftarrow$ \textbf{Update}($\hat{x}_{k|k-1},P_{k|k-1},U_k,\boldsymbol{A},\boldsymbol{M}$)

\State \Return $\hat{x}_k, P_k$

\end{algorithmic}
\end{algorithm}

\subsection{Prediction step}
The prediction step involves an integration of the gyroscopic input, as described in detail in~\cite{Maley2013MultiplicativeQE}.

\begin{algorithm}
\caption{MEKF Prediction Step}\label{alg:Prediction}
\begin{algorithmic}[1]

	\Require Previous state $\hat{x}_{k-1}$, covariance $P_{k-1}$, process noise $Q_k$, gyroscope $\boldsymbol{\omega} = (p, q, r)$, time step $dt$
	\Ensure Updated state $\hat{x}_{k|k-1}$, covariance $P_k$

	\State $[\omega_x, \omega_y, \omega_z] \gets [p - B_{k,x}, q - B_{k,y}, r - B_{k,z}]$
    \State $\Delta q \gets \exp([\omega_x, \omega_y, \omega_z] \cdot dt)$
	\State $q_{k|k-1} \gets \exp(\hat{x}[1:3]) \cdot \Delta q$ 
	\State $\hat{x}[1:3] \gets \log(q_{k|k-1})$
    \State $\hat{x}[4:6] \gets \hat{x}[4:6]$
    \State Initialize $F \gets 0_{6 \times 6}$, $\Phi \gets I_6$
    \State $F_{1:3,1:3} \gets -\text{skew}([\omega_x, \omega_y, \omega_z])$
    \State $F_{1:3,4:6} \gets -I_3$
    \State $\Phi \gets I_6 + F \cdot dt$ \Comment{For large $dt$, higher-order terms may be required}
	\State $P_k \gets \Phi \cdot P_{k-1} \cdot \Phi^\top + Q_k$
	\State \Return $\hat{x}_{k|k-1}, P_k$
\end{algorithmic}
\end{algorithm}

Given gyroscopic data $[\omega_x, \omega_y, \omega_z]$, the integration computes
	\begin{equation}
	q_{k|k-1} = q_{k-1} \cdot \exp([\omega_x, \omega_y, \omega_z] \cdot dt),
	\end{equation}
so that $\log(q_{k|k-1})$ corresponds to $\hat{x}[1:3]$ before the update step.

The matrix $F$ computed in Algorithm~\ref{alg:Prediction} is a $6 \times 6$ matrix such that
	\begin{equation}
\dot{x} \approx (I_6 + F) \cdot x.
	\end{equation}
It is derived from the differential equation satisfied by the error vector $\eta$, where for a full quaternion $q$ and its estimate $\hat{q}$:
	\begin{equation}
q = \hat{q} \cdot \begin{pmatrix} 1 \\ \eta/2 \end{pmatrix},
	\end{equation}
and
	\begin{equation}
\dot{\eta} = -[\omega]_\times \eta - \beta_\omega - \zeta_\omega,
	\end{equation}
with $\beta_\omega$ the gyroscope bias and $\zeta_\omega$ the gyroscope noise.

The solution of this equation is approximated by the exponential
	\begin{equation}
\exp(-[\omega]_\times \cdot dt) \approx I_3 - [\omega]_\times dt.
	\end{equation}
The covariance is propagated using the linearization $\Phi$, which is a first-order approximation. A higher-order approximation can be used:
	\begin{equation}
\Phi = I_6 + F \cdot dt + \frac{1}{2} F^2 \cdot dt^2,
	\end{equation}
as described in~\cite{Maley2013MultiplicativeQE}. This is particularly important for synthetic data, where the first-order approximation performs poorly under large gyroscope variations.

\subsection{Update Step}

The update step is described in Algorithm~\ref{alg:Update}. It combines the inner state predicted in the previous step with the state inferred from accelerometer and magnetometer measurements. This fusion inherits favorable asymptotic properties from the optimality of the Kalman filter, as discussed in~\cite{10.1115/1.3662552}: for linear dynamics with Gaussian noise, it provides the optimal minimum-variance estimate.

In the Multiplicative Extended Kalman Filter (MEKF) variant, the inner vector state is interpreted as a perturbation $\delta q$ of the quaternion representing the orientation. Accordingly, the update step is applied as
\[
q_k \gets q_{k|k-1} \cdot \exp(\delta q).
\]

\begin{algorithm}
\caption{MEKF Update Step}\label{alg:Update}
\begin{algorithmic}[1]

    \Require Predicted state $\hat{x}_{k|k-1}$, Covariance $P_{k|k-1}$, Measurement noise $U_k$, Accelerometer $\boldsymbol{A}$, Magnetometer $\boldsymbol{M}$
    \Ensure Updated state $\hat{x}_k$, Covariance $P_k$

    \State $z_k \gets (\boldsymbol{A}, \ \boldsymbol{M})$
    \State $y_k \gets z_k - H_k \hat{x}_{k|k-1}$ \Comment{Innovation}
    \State $S_k \gets H_k P_{k|k-1} H_k^\top + U_k$ \Comment{Innovation covariance}
    \State $K_k \gets P_{k|k-1} H_k^\top S_k^{-1}$ \Comment{Kalman gain}
    \State ${\delta q}_k \gets K_k y_k$ \Comment{Quaternion perturbation}
    \State $q_k \gets q_{k|k-1} \cdot \exp({\delta q}_k)$
    \State $\hat{x}_k \gets \log(q_k)$
    \State $P_k \gets (I - K_k H_k) P_{k|k-1}$

    \State \Return $\hat{x}_k, P_k$

\end{algorithmic}
\end{algorithm}

In this algorithm, the measurement noise covariance $U_k$ and the transition matrix $H_k$ must be defined.  

The noise covariance is often chosen to be diagonal:
\[
U_k = \mathrm{diag}(u_0,u_1,u_2,u_3,u_4,u_5),
\] 
where the $u_i$ are tuning parameters of the filter.  

The transition matrix $H_k$ can be expressed in $3\times 3$ blocks:
\[
H_k = \begin{bmatrix}
    [R(q)\boldsymbol{g}]_{\times} & 0 \\
	[R(q)\boldsymbol{b}]_{\times} & 0
\end{bmatrix},
\]
where $R(q)$ is the rotation matrix associated with quaternion $q$, $\boldsymbol{g}$ is the gravity vector, and $\boldsymbol{b}$ is the magnetic field vector in the Earth frame.

The derivation of this transition matrix is discussed in~\cite{Maley2013MultiplicativeQE}:
\begin{equation}
\label{eq:derivation_measure0}
\mathbf{m}^\mathcal{R} = R(q)^{-1} \mathbf{m}^\mathcal{G} = R(\hat{q}\cdot \exp(\delta q))^{-1}\mathbf{m}^\mathcal{G},
\end{equation}
for a measurement $\mathbf{m}$, which leads to
\begin{equation}
\label{eq:derivation_measure1}
\mathbf{m}^\mathcal{R} - \hat{\mathbf{m}}^\mathcal{R} \approx [I-[\delta q]_\times]R(q)^{-1} \mathbf{m}^\mathcal{G}.
\end{equation}

The relation above arises from the following approximation for a quaternion $q$:
\begin{equation}
R(\delta q) \approx I-2[q_{2:4}]_\times,
\end{equation}
which implies
\begin{equation}
R(q) \approx R(\hat{q}) \cdot (I+[\delta q]_\times).
\end{equation}

Thus,
\begin{equation}
\delta \mathbf{m}^\mathcal{R} = \begin{pmatrix} [R(q)\cdot \mathbf{m}^\mathcal{G}]_\times & 0_{3\times 3} \end{pmatrix} \cdot \begin{pmatrix}\delta q \\ 0\\\end{pmatrix}.
\end{equation}

Similarly, the accelerometer measurement $\mathbf{A}$ can be related to the gravity vector $\Vec{g} = [0, 0, g]^\top$ as
\begin{equation}
\mathbf{A} -\hat{\mathbf{A}} = -[\delta q]_\times R(q) \Vec{g}.
\end{equation}

This expression directly leads to the transition matrix $H$ in the form
\begin{equation}
H = \begin{pmatrix}
    [R(q) \Vec{g}]_\times & 0\\
    [R(q) \Vec{b}]_\times & 0\\
\end{pmatrix}.
\end{equation}

\subsection{Multiplicative Residual}

Observing that the transition matrix corresponds to the derivative of
$\mathbf{m}^\mathcal{R} - \hat{\mathbf{m}}^\mathcal{R}$, as given in Equation~\ref{eq:derivation_measure0}, with respect to $\delta q$, which leads to Equation~\ref{eq:derivation_measure1}, 
we can generalize this approach to another error function, referred to as the \emph{residual}.
In Equation~\ref{eq:derivation_measure0}, the residual $y_k$ is defined as
\begin{equation}
y_k \gets z_k - H_k \hat{x}_{k|k-1}.
\end{equation}

A \emph{multiplicative residual} can be defined as
\begin{equation}
\mathbf{A} \times \hat{\mathbf{A}}
\end{equation}
for the acceleration, as described in~\cite{zanetti:mult_residual}.
Applying the same formulation to the magnetometer, we obtain the following transition matrix $H$:
\begin{equation}
H = \begin{pmatrix}
    I - (R(q) \Vec{g}) (R(q) \Vec{g})^T & 0 \\
    I - (R(q) \Vec{b}) (R(q) \Vec{b})^T & 0
\end{pmatrix}.
\end{equation}

\section{Reversible Filter}
\label{sec:revmekf}

We introduce the key concepts behind the main algorithm proposed in~\cite{covanov2025reversiblekalmanfilterstate}, and in the following sections, we propose a classifier that adapts it to practical situations.

\subsection{Theoretical Aspect}

In this section, we define the mathematical framework useful for comparing different filters. The core idea is that for certain actions on measurements, one can define a \emph{reversible filter}, which can be interpreted as a filter that allows backward propagation.

This concept originates from the observation that, for example, in an IMU, measurements such as the gravity vector and magnetometer should reflect transformations due to rotations.

Formally, let $G$ be a group, which can be considered as $SO(3)$. Let $\mathcal{M} = G \times V$ denote the set of measurements, where $V$ is a vector space on which $G$ acts. We define a left action of $G$ on $\mathcal{M}$ by
\[
h * (l,w) = (h \cdot l, h \cdot w),
\]
for $(l,w) \in \mathcal{M}$.

A filter is a function $f$ mapping $(\mathcal{S},\mathcal{M})$ to $\mathcal{S}$, where $\mathcal{S}$ is the set of state vectors. We define the \emph{$G$-reversibility} property as
\[
\forall u, \forall m = (h,w), \quad f(f(u,m), h^{-1} * (Id_G, w)) = u.
\]

This property can be strengthened to account for sensor errors. A \textbf{strong} \emph{$G$-reversibility} property for a filter $f$ is defined as
\[
\forall \mathbf{\epsilon}, \forall u, \forall m, \quad f(f(u,m), h^{-1}*(Id_G, w)+\mathbf{\epsilon}) = u + O_f(\mathbf{\epsilon}),
\]
meaning that a measurement error of magnitude $\epsilon$ leads to a bounded error of magnitude $O_f(\epsilon)$, depending only on the function $f$. This property allows achieving arbitrary precision. However, classical variants of the MEKF do not satisfy it.

\subsection{MEKF is Equivariant but Not Reversible}

Here, we show that the Multiplicative Extended Kalman Filter (MEKF) is not reversible.  

A counterexample uses the accelerometer, which, when affected by external acceleration, does not exactly reflect a rotation of the gravity vector. Assume noiseless IMU sensors at time $t=0$, with no initial rotation, so measurements are of the form
\[
m = (Id_{SO(3)}, (\mathbf{A}, \mathbf{M})).
\]

Consider $\mathbf{M} = [1,0,0]$ and $\mathbf{A} = [0,0,1] + y \cdot [0,1,0]$, where $y \cdot [0,1,0]$ represents external acceleration along the $y$ axis. For small $y$, the MEKF correction yields
\[
f([0,0,0], m) \approx [0,0,0],
\]
meaning the axis-angle representation of the orientation remains near zero. For large $y$, the behavior is 
\[
	f([0,0,0], m) \approx_{x\rightarrow +\infty} \left[\frac{\pi}{2},0,0\right],
\]
indicating the axis-angle representation of the estimated orientation is closer to $[\frac{\pi}{2},0,0]$, which is a rotation around $X$-axis of angle $\frac{\pi}{2}$.

Applying the reversibility property would require evaluating
\[
f(f([0,0,0], m), (Id_{SO(3)}, (\mathbf{A}, \mathbf{M}))),
\]
which produces a quaternion logarithm close to $[\frac{\pi}{2},0,0] \neq [0,0,0]$.  

Reversibility is orthogonal to the equivariance property discussed in~\cite{FORNASIER2025112495}, which captures MEKF symmetries. In our approach, we propose a MEKF variant that compensates for asymmetry by leveraging surface knowledge and accelerometer data.

\subsection{Geometric Constraints Solving}

We introduce an intermediate step between the prediction and update steps that exploits knowledge of the plane normal on which the IMU moves. By incorporating this normal, one can derive the gravity vector from input data and replace raw accelerometer measurements during the MEKF update.

Given IMU measurements $(\mathbf{\omega}, \mathbf{A}, \mathbf{M}, \Vec{n})$, we decompose them as
\begin{equation}
m = (\mathbf{\omega}, \mathbf{A}_g^R + \mathbf{A}_{\text{ext}}^R, \mathbf{M}, \Vec{n}),
\end{equation}
where $\mathbf{A}_g$ is the rotated gravity vector.

For planar motion, let $\mathbf{p}_k$ and $\mathbf{v}_k$ denote position and velocity at step $k$. Then
\begin{equation}
\label{eq:positionacc}
\mathbf{p}_{k} = \mathbf{p}_{k-1} + \mathbf{v}_{k} \cdot dt.
\end{equation}
Since $p_{k-1}\cdot \Vec{n} = p_{k}\cdot \Vec{n} = 0$, we have:
\begin{equation}
\label{eq:speedacc}
	\mathbf{v}_{k}\cdot \Vec{n} = (\mathbf{v}_{k-1} - \Vec{g} \cdot dt + R(q_k) \cdot \mathbf{A}_k^\mathcal{R} \cdot dt)\cdot \Vec{n} = 0.
\end{equation}
One can inject in Equation~\ref{eq:speedacc} the following relation
\begin{equation}
\mathbf{A}_{\text{ext}}^\mathcal{G} = \Vec{g} + R(q_k) \cdot \mathbf{A}_k^\mathcal{R}.
\end{equation}
Thus, there exists a rotation $R$ for which the measurement can be rewritten as
\begin{equation}
m = (\mathbf{\omega}, R^{-1} \cdot \Vec{g} + R^{-1} \cdot \mathbf{A}_{\text{ext}}^\mathcal{G}, R \cdot \Vec{b}, \Vec{n}).
\end{equation}

Given the plane constraint $\mathbf{v}_k \cdot \Vec{n} = 0$ and $\mathbf{v}_{k-1} \cdot \Vec{n} = 0$, there are finitely many rotations $R$ satisfying
\begin{equation}
(- \Vec{g} \cdot dt + R \cdot \mathbf{A}_k^\mathcal{R} \cdot dt) \cdot \Vec{n} = 0
\end{equation}
and
\begin{equation}
R \mathbf{M}_k^\mathcal{R} = \Vec{b}.
\end{equation}

\begin{figure}[h!]
  \centering
  \includegraphics[width=0.33\textwidth]{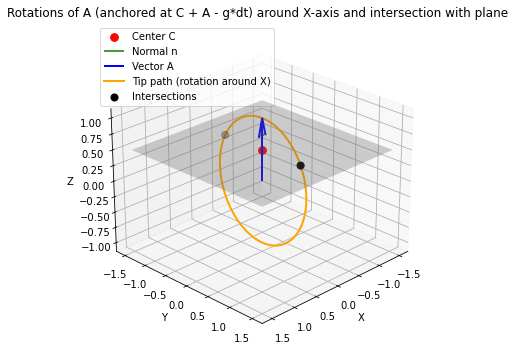}
  \caption{Example of rotations of acceleration around the $X$ axis and their intersection with a plane.}
  \label{fig:rotation_acc}
\end{figure}

Figure~\ref{fig:rotation_acc} illustrates a scenario where $\Vec{b}$ is colinear with the $X$ axis and the plane normal is colinear with the $Z$ axis. The center $C$ is the projection of $\mathbf{v}_{k-1}$ onto the plane.

\begin{definition}[Distance Function]
\label{def:distance}
For each sample $k$, assuming $\|\mathbf{M}_k^\mathcal{R}\|=1$, let $R' = R(\mathbf{M}_k^\mathcal{R} \to \Vec{b})$ be the unique rotation along $\mathbf{M}_k^\mathcal{R} \otimes \Vec{b}$ such that $R'(\mathcal{M}_k^\mathcal{R}) = \Vec{b}$. Denote the corresponding unit quaternion as $q'$. Let
\[
\theta_0 = \arg \min_\theta \|\log(q_{k|k-1} \cdot q'^{-1}) - \theta \Vec{b}\|_2.
\]
Define the distance function $h_k(\theta)$ as
\[
\theta \mapsto (- \Vec{g} \cdot dt + \exp((\theta-\theta_0)\Vec{b}) \cdot R' \cdot \mathbf{A}_k^\mathcal{R} \cdot dt) \cdot \Vec{n}.
\]
If $h_k(\theta) = 0$, then $\exp((\theta+\theta_0)\Vec{b}) \cdot q'$ yields a rotation $R$ such that
\[
(- \Vec{g} \cdot dt + R \cdot \mathbf{A}_k^\mathcal{R} \cdot dt) \cdot \Vec{n} = 0.
\]
\end{definition}

\begin{remark}[Uniqueness of Angle]
\label{rem:unicity}
Solving $h_k(\theta) = 0$ typically yields two solutions, $\theta_1$ and $\theta_2$. Choosing the one closest to $\theta_0$ (the gyroscope prediction) ensures uniqueness in generic cases.
\end{remark}
\begin{remark}[Non planar movement]
	One can adapt this approach to a non linear trajectory such that a normal $\Vec{n}_k$ is given at each sample, but it requires to add the speed to the state vector and to use the relation:
\begin{equation}
	(v_{k-1}- \Vec{g} \cdot dt + R \cdot \mathbf{A}_k^\mathcal{R} \cdot dt) \cdot \Vec{n} = 0
\end{equation}
	since the relation ${v}_{k-1} \cdot \Vec{n}_k=0$ is not true anymore.

	Our approach is to make the following approximation:
\begin{equation}
	\label{eq:approxnonplanar}
	{v}_{k-1} \cdot \Vec{n}_k \approx 0.
\end{equation}

\end{remark}

By combining these constraints, we can compute a unique rotation closest to the MEKF prediction, from which we deduce corrected measurements $(\mathbf{\omega}, \mathbf{A}_g^R, \mathbf{M}, \Vec{n})$.

\subsection{Pseudo-code}

Adding this intermediate step results in a reversible variant of the Kalman filter (Algorithm~\ref{alg:RevMEKF}). The reversibility arises because:
\begin{itemize}
    \item Additional information is provided by the known plane normal.
    \item Each measurement term depends only on orientation. Given $R$ such that $R \cdot \Vec{g} = \mathbf{A}$ and $R \cdot \Vec{b} = \mathbf{M}$,
    \[
    h^{-1}*(Id,(\mathbf{A},\mathbf{M})) = (h^{-1}, (h^{-1} R \cdot \Vec{g}, h^{-1} R \cdot \Vec{b}))
    \]
    and
    \[
    f(f(u, (h, (\mathbf{A}, \mathbf{M}))), h^{-1} * (Id_{SO(3)}, (\mathbf{A}, \mathbf{M}))) = u.
    \]
\end{itemize}

This property, combined with the continuity of the linear algebra operations, guarantees strong reversibility.

\begin{algorithm}
\caption{Reversible Kalman Filter: Prediction and Update}\label{alg:RevMEKF}
\begin{algorithmic}[1]
\Require Previous state estimate $\hat{x}_{k-1}$, Covariance $P_{k-1}$, Gyroscope $\mathbf{\omega} = (p,q,r)$, Accelerometer $\mathbf{A}$, Magnetometer $\mathbf{M}$, Plane normal $\Vec{\mathbf{n}}$, Time step $dt$
\Ensure Updated state estimate $\hat{x}_k$, Covariance $P_k$

\Statex $\hat{x}_{k|k-1}, P_{k|k-1} \leftarrow$ \textbf{Prediction}($\hat{x}_{k-1}, P_{k-1}, Q_k, \mathbf{\omega}, dt$)
\Statex $\mathbf{A}_g \leftarrow$ \textbf{GeoSolving}($\hat{x}_{k|k-1}, \mathbf{A}, \mathbf{M}, \Vec{\mathbf{n}}$)
\Statex $\hat{x}_k, P_k \leftarrow$ \textbf{Update}($\hat{x}_{k|k-1}, P_{k|k-1}, U_k, \mathbf{A}_g, \mathbf{M}$)

\State \Return $\hat{x}_k, P_k$
\end{algorithmic}
\end{algorithm}

\subsection{Behaviour on Synthetic Data}
\label{sec:synthdata}

Synthetic data are useful for comparing the expected behavior of MEKF and Rev-MEKF.

First, one needs to define a metric to compare different approaches.
\begin{definition}[Metric function]
	\label{def:metric}
	Given a filter whose output is a serie of vector states $X = (\hat{x}_k)_k$ such that
	$\hat{x}_k[1:3]$ is the logarithm of a rotation,
	we compare it to another serie of vector states $Y = (\hat{y}_k)_k$ by using the following function:
	$$\Lambda(X,Y) = \sum_k |1-Re(\exp(x_k[1:3])\cdot \exp(-y_k[1:3]))|$$
	where $Re(q)$ is the real part of quaternion $q$.
\end{definition}
Definition~\ref{def:metric} can be thought as the cumulated difference of two series of vectors and is related to the chordal distance between two vectors. Thus, given a ground truth $\mathcal{T} = (t_k)_k$, for a serie of estimates $X$, we use the estimate given by
$$\Lambda(X,\mathcal{T}).$$
One can verifiy that this metric is equal to $0$ if and only if $X=\mathcal{T}$.

The data used to analyze typical MEKF behavior were generated along a $2D$ random trajectory on a plane with a normal vector $[0,-1,1]$, corresponding to $100$ samples, without introducing any noise. Curves were produced for different levels of numerical precision using the multiprecision float library \texttt{mpf}.

We considered three experimental conditions: 
\begin{enumerate}
    \item no noise, 
    \item gyroscope bias of order $10^{-15}$ $\text{rad.s}^{-2}$ with no accelerometer noise, 
    \item accelerometer with varying noise levels, and gyroscope noise of order $10^{-15}$ $\text{rad.s}^{-2}$.
\end{enumerate}

\begin{figure}[h!]
  \centering
  \includegraphics[width=0.9\columnwidth]{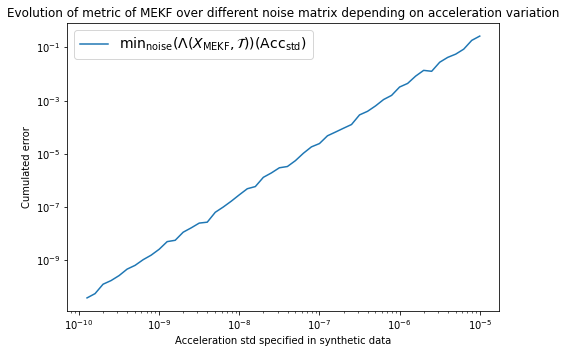}
  \caption{$\max(\Lambda(X_{\text{MEKF}},\mathcal{T}))$ as a function of average acceleration variations during $dt=1/100s$.}
  \label{fig:evolution_pos}
\end{figure}

This allows us to quantify the intrinsic loss of the algorithm by studying numerical imprecision. In Figure~\ref{fig:evolution_pos}, we analyze the maximal drift of the metric $\Lambda(X_{\text{MEKF}},\mathcal{T})$ with noiseless sensors. The precision depends on the dynamics of the motion rather than solely on the quality of the data. 

\begin{figure}[h!]
  \centering
  \includegraphics[width=0.9\columnwidth]{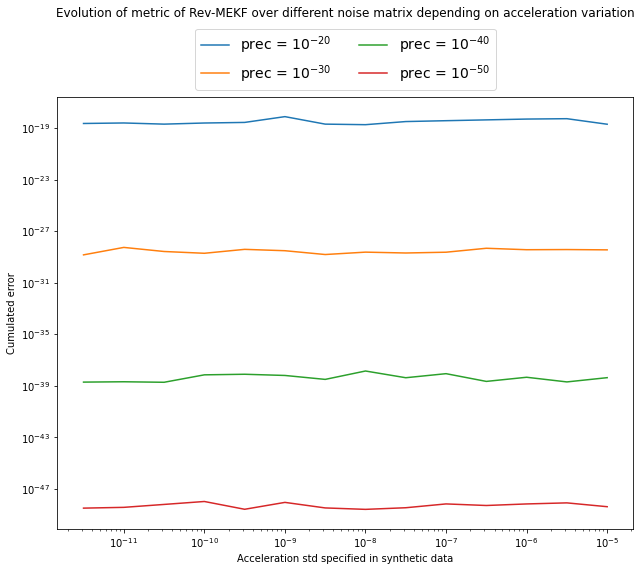}
  \caption{$\max(\Lambda(X_{\text{Rev-MEKF}},\mathcal{T}))$ as a function of average acceleration variations during $dt=1/100s$.}
  \label{fig:revmekf_vs_accvar}
\end{figure}

From this, we conclude that MEKF's orientation estimate is intrinsically sensitive to accelerometer variations: smaller variations yield more accurate estimates. In contrast, Rev-MEKF provides an estimate for which the metric has an order of magnitude equal to $N \cdot \epsilon$, where $N$ is the number of samples and $\epsilon$ is the base precision ($10^{-20}$, $10^{-30}$, $10^{-40}$, or $10^{-50}$), as shown in Figure~\ref{fig:revmekf_vs_accvar}. Applied to synthetic data, Rev-MEKF’s numerical precision is independent of the motion type, making it suitable for testing MEKF’s update step implementation. With noiseless sensors, it can generate orientation with arbitrary precision.

\begin{figure}[h!]
  \centering
  \includegraphics[width=0.9\columnwidth]{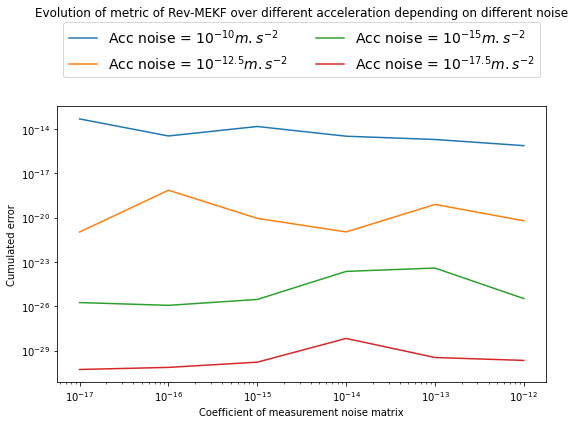}
  \caption{$\max(\Lambda(X_{\text{Rev-MEKF}},\mathcal{T}))$ as a function of measurement noise matrix coefficients.}
  \label{fig:revmekf_noise_vs_accvar}
\end{figure}

One can study the behavior of Rev-MEKF by introducing acceleration noise. In order to do that, we fix the precision to $10^{-50}$ and generate acceleration noise of $10^{-10}$, $10^{-12.5}$, $10^{-15}$, or $10^{-15}$ $\text{m.s}^{-2}$, along with gyroscope noise of order $10^{-15}$ $\text{rad.s}^{-1}$, as shown in Figure~\ref{fig:revmekf_noise_vs_accvar}, we see that Rev-MEKF’s error is bounded either by numerical precision or by accelerometer accuracy.
This figure shows that in a noisy context, when changing the acceleration variation, the order of magnitude of the final metric stays inferior to the noise of the acceleration.

Considering a gyroscope bias of $10^{-15}$ $\text{rad.s}^{-1}$, we observe that the cumulated error has some variations.
This phenomena can be explained by the fact that the update step has an intrinsic imprecision due to convergence that can be measured in a context of a static motion and with an isolated perturbation. Thus, the error of Rev-MEKF (measured by our metric) is bounded only by sensor noise or the intrinsic imprecision of the update step. With a biased gyroscope, Rev-MEKF converges quickly to the true orientation, but final error depends on the MEKF update step’s convergence properties.

\section{Analysis of Heuristic Rev-MEKF on static Real Data}

\label{sec:evalmekf}

We provide an analysis of the behavior of the Rev-MEKF on real data collected from an ellipse-D SBG device. Real-world data introduce additional challenges, which we discuss in this section.% The dataset used comes from~\cite{jeferson_menegazzo_2021}.

\subsection{Raw Rev-MEKF}

We conducted experiments under several scenarios using different sensor configurations, including odometry and IMU sensors. We describe here only the behavior on an IMU sensor, which allows already to understand the limit of the theoretical approach. The main observation is that the Rev-MEKF performs well under idealized conditions. However, its performance in real-world scenarios is affected by several factors:
\begin{itemize}
\item noise in the accelerometer and magnetometer measurements;
\item errors in the computation of the gravity and magnetic field vectors;
\item magnetic disturbances caused by electronic devices.
\end{itemize}

As a result, the average distance between the finite set of rotations corresponding to the intersection of the acceleration vector with the surface (as defined in the previous section) becomes smaller than the noise induced by these factors. Consequently, distinguishing between two intersection points, as in Figure~\ref{fig:rotation_acc}, becomes impossible.

Our expectation is that this average distance increases when the acceleration deviates from the gravity vector, i.e., during strong external accelerations, which correspond to situations where the MEKF performs poorly.

\begin{figure}[h!]
\centering
\includegraphics[width=0.45\columnwidth]{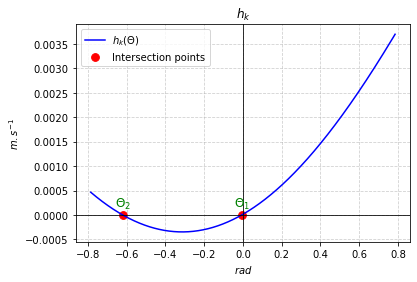}
\includegraphics[width=0.45\columnwidth]{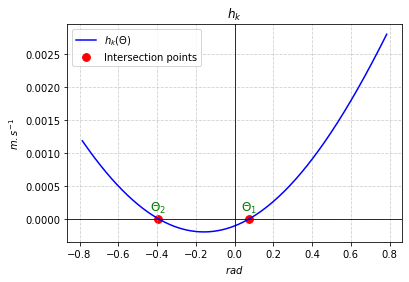}
\caption{Typical $h_k$ with $\text{std}(\mathbf{A}^\mathcal{R})/\text{freq}$ equal to $10^{-3}$ on the right and $10^{-5}$ on the left.}
\label{fig:comparison_distance_hk_synth}
\end{figure}

To estimate whether the behavior at a given sample $k$ is good or bad, one must consider the function $h_k$ defined in Definition~\ref{def:distance}.
Figure~\ref{fig:comparison_distance_hk_synth} shows a typical evolution of the function $h_k$ on synthetic data along a random trajectory.

We observe that for large variations of the acceleration, there are clearly two intersection points, whereas this distance becomes smaller as the variation tends to zero.

\begin{figure}[h!]
\centering
\includegraphics[width=0.6\columnwidth]{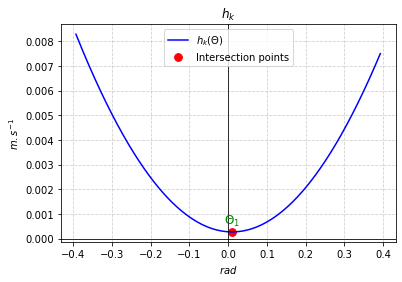}
	\caption{Typical $h_k$ with a static SBG device where $|\theta| < \frac{\pi}{8}$: the curve is not intersecting the line $y=0$.}
\label{fig:comparison_distance_hk_real}
\end{figure}

During static phases, the behavior of the Rev-MEKF, as defined in Section~\ref{sec:revmekf}, is the worst.
We tested this behavior with an IMU (Ellipse-D from SBG), as shown in Figure~\ref{fig:comparison_distance_hk_real}.

In this situation, the noise on the acceleration norm prevents the algorithm from finding a valid solution, leading to errors. Moreover, the curve $h_k$ is very flat around $\theta \approx 0$, which makes it difficult to identify the correct angle.

\begin{figure}[h!]
\centering
\includegraphics[width=0.75\columnwidth]{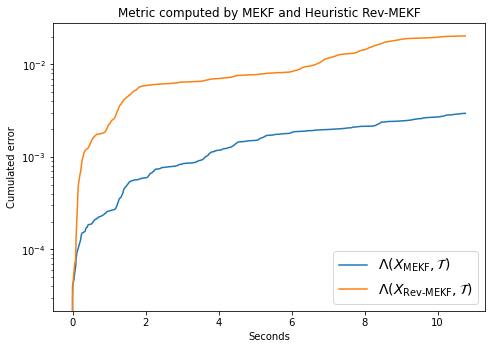}
	\caption{$\Lambda(X_\text{MEKF},\mathcal{T})$ and $\Lambda(X_\text{Rev-MEKF},\mathcal{T})$ with a static IMU (SBG device).}
\label{fig:static_revmekfdrift}
\end{figure}

Under these conditions, using $Q_k = 10^{-2}\cdot I_{6\times 6}$ and $U_k = 10^{-1}\cdot I_{6\times 6}$, the Rev-MEKF drifts, as illustrated in Figure~\ref{fig:static_revmekfdrift}.
The ground truth $\mathcal{T}$ is set to be the constant quaternion equal to the mean of the quaternions obtained directly from the accelerometer and the magnetometer.

Thus, during static phases with noisy sensors, the Rev-MEKF performs worse than the MEKF because the rotation angle $\alpha$, which projects the acceleration onto its closest point on the plane, can become large due to errors in the gravity vector and sensor noise.

\subsection{Heuristic Solution}

To compensate for the problem observed above, a decision algorithm is required to switch between different situations detected by the distance function $h_k$ defined in Definition~\ref{def:distance}.

The Rev-MEKF can be adapted to apply geometrically constrained corrections at appropriate times. To this end, the first approach developed in~\cite{covanov2025reversiblekalmanfilterstate} introduces a detection step that determines when to switch between MEKF and Rev-MEKF, based on a trade-off between sensor noise and the type of motion.
To mitigate the issue described in the previous section, one must select when the prediction step falls within the class of rotations where the intersection point used in Rev-MEKF is most likely.

We use the following modifications, which is close to the heuristic used in~\cite{covanov2025reversiblekalmanfilterstate}:
\begin{itemize}
\item To avoid errors due to global magnetic field computation, use
    \[
	    q_{k|k-1}^{-1} \cdot [0,1,0,0] \cdot q_{k|k-1}
    \]
instead of $q^{-1} \cdot \Vec{b} \cdot q$ on the raw magnetometer data, where $q_{k|k-1}$ is the predicted quaternion at sample $k$.
This may introduce a small error due to the prediction step on the vector $\Vec{b}$, but it removes dependency on the precision of $\Vec{b}$ computation.
\item By default, return the vector normally retuned by MEKF.
\item If the distance from the predicted rotation to the closest intersection rotation is $\gamma$ times smaller than the distance to the default rotation, return the vector normally used by the Rev-MEKF.
\end{itemize}

\begin{figure}[h!]
\centering
\includegraphics[width=0.75\columnwidth]{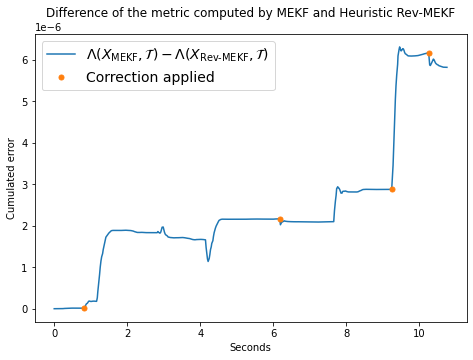}
	\caption{Difference $\Lambda(X_\text{MEKF},\mathcal{T})-\Lambda(X_\text{Rev-MEKF},\mathcal{T})$ with a static IMU (SBG device) and $\gamma = 1$.}
\label{fig:static_heuristicrevmekf}
\end{figure}

With this heuristic, the behavior observed in the left plot of Figure~\ref{fig:comparison_distance_hk_synth} is expected. In practice, the returned acceleration is significantly less noisy.
Figure~\ref{fig:static_heuristicrevmekf} shows that the Heuristic Rev-MEKF drifts less than the MEKF with this modification since the difference 
\[
\Lambda(X_\text{MEKF},\mathcal{T})-\Lambda(X_\text{Rev-MEKF},\mathcal{T})
\]
is positive. We already can observe how the metric defined in Definition~\ref{def:metric} can compare the behavior of $2$ different variants.

\begin{figure}[h!]
\centering
\includegraphics[width=0.33\textwidth]{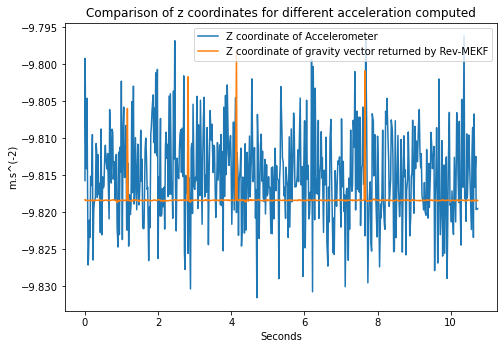}
\caption{Comparison of the $z$-coordinates of the gravity vector returned by the geometrically constrained correction step with the original accelerometer $z$-coordinates.}
\label{fig:comparison_acc_dataset}
\end{figure}

The most notable improvement is observed in the gravity vector transmitted to the update step, shown in Figure~\ref{fig:comparison_acc_dataset}, which demonstrates a clear reduction in noise.

\begin{figure}[h!]
\centering
\includegraphics[width=0.7\columnwidth]{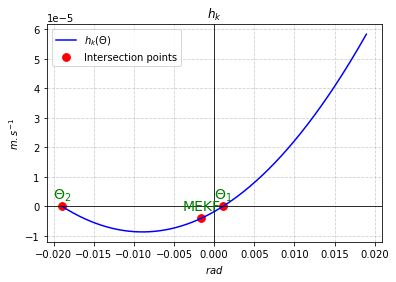}
\caption{Curve $h_k$ for typical cases of corrections with the heuristic.}
\label{fig:typical_correction}
\end{figure}

It is important to carefully select $\gamma$ based on the application, plane normal, and typical accelerations. A value of $\gamma = 1$ allows discrimination between situations where the probability that the best point is returned by the MEKF is lower than that returned by the Rev-MEKF.
Figure~\ref{fig:typical_correction} shows typical cases of applied corrections.% In some instances, corrections that might be expected are not triggered by the heuristic.

However, in static conditions, we can't distinguish the improvement that Heuristic Rev-MEKF can bring and a random event detection. This is why we provide a comparison of its behavior on several datasets, in different type of conditions, to prove that the heuristic that has been chosen has a physical meaning.

\section{Comparison on different datasets}
\label{sec:datasets}
In this section, we apply the previous heuristic version to several datasets, for which there is a ground truth recorded.
The main goal is to show that the heuristic has a physical meaning, which validates its use in particular situations.
We use the metric defined in Definition~\ref{def:metric} to show that moments where Rev-MEKF improves on MEKF correspond to characteristic points of its curve.
We used the following datasets:
\begin{itemize}
	\item a serie of tests of inertial and magnetic sensors using an industrial robot as ground-truth developed in~\cite{LowcomplexityMARG};
	\item the OxIOD dataset~\cite{chen2018oxioddatasetdeepinertial} recording human motion during walking and running;
	\item the INSANE dataset~\cite{brommer_insane_2022} recording the motion of a UAV in conditions close to Mars environment;
	\item the underwater caves sonar dataset~\cite{10.1177/0278364917732838} recording the motion of an autonomous underwater vehicle onn an almost planar surface.
\end{itemize}
The OxIOD and the underwater caves sonar datasets~\cite{chen2018oxioddatasetdeepinertial,10.1177/0278364917732838} have already been used in DANAE~\cite{russo2021danae++}, a software allowing one to denoise attitude estimation using deep learning approach. We have been able to reuse part of the code to compare ground-truth to our approach.

\subsection{Global statistics on different datasets and quantitative analysis}

In this section, we compare the effect of Rev-MEKF on the different datasets mentioned previously in a quantitative way: we take each file and split it into time intervals of various lengths (in seconds) and
we use as an indicator of improvement the sequence defined by
\[
    \Delta : k \mapsto \Lambda(X_{\text{MEKF}},\mathcal{T})_k - \Lambda(X_{\text{Rev-MEKF}},\mathcal{T})_k.
\]
When
\begin{equation}
    \label{eq:reldelta}
    \Delta(k) > \Delta(k-1),
\end{equation}
Rev-MEKF approaches the ground truth better than MEKF. Thus, for each interval $(t_j)$, we compute the proportion of samples $k$ for which Equation~\ref{eq:reldelta} holds, which correspond to moments when Rev-MEKF is drifting slower from Ground Truth than MEKF.

In Table~\ref{tab:compare_datasets}, we use this score to analyze the proportion of positive Rev-MEKF performance values over time intervals for each file.  The Ground truth that is used in this table is directly extracted from files associated to the datasets given in~\cite{LowcomplexityMARG,chen2018oxioddatasetdeepinertial,10.1177/0278364917732838,brommer_insane_2022} and for which
an existent repository specifies how to obtain a quaternion.
\newcolumntype{P}[1]{>{\centering\arraybackslash}p{#1}}
\begin{center}
	\begin{table}[h!]
\renewcommand{\arraystretch}{1.2}
\resizebox{\columnwidth}{!}{
		%{
{\Huge
%\newcolumntype{C}{>{\centering\arraybackslash}m{10cm}}
\begin{tabular}{|l|P{8.5cm}|P{8.5cm}|P{8.5cm}|P{8.5cm}|}
%\begin{tabular}{|m{3.5cm}|C|C|C|C|}
\hline

 \multicolumn{5}{|l|}{
		\textbf{File:} V1000/Sequencia1\quad
	    \textbf{Raw data:} MPU9150\quad
		\textbf{Ground truth (same as{~\cite{LowcomplexityMARG}}):} Xsens MTi-30 \quad
    } \\
\hline

\textbf{Parameters}
  & \multicolumn{4}{c|}{$Q_k=10^{-2}I_6$, $U_k=10^{2}I_6$, $\gamma = 1.0$} \\
\hline

\textbf{Intervals} 
  & $0\le t<2.5s$ & $2.5s\le t<5s$ & $5s\le t<7.5s$ & $7.5s\le t<10s$ \\
\hline

\textbf{Percentages} 
  & 41\% & 46\% & 57\% & 0\% \\
\hline

\textbf{Intervals} 
  & $9s\le t<12s$ & $12s\le t<15s$ & $15s\le t<18s$ & \textbf{Total} \\
\hline

\textbf{Percentages} 
  & 12\% & 53\% & 65\% & \textbf{36\%} \\
\hline

\multicolumn{5}{|l|}{
		\textbf{File:} Caves sonar dataset\quad
	    \textbf{Raw data:} ADIS16480\quad
		\textbf{Ground truth (same as{~\cite{russo2021danae++}}):} ADIS16480\quad
    } \\
\hline
\textbf{Parameters}
  & \multicolumn{4}{c|}{$Q_k=10^{-2}I_6$, $U_k=I_6$, $\gamma = 1.0$} \\
\hline

\textbf{Intervals} 
  & $0\le t<275s$ & $275\le t<550s$ & $550s\le t<825s$ & $825s\le t<1100s$ \\
\hline

\textbf{Percentages} 
  & 50\% & 58\% & 44\% & 42\% \\
\hline

\textbf{Intervals} 
  & $1100s\le t<1375s$ & $1375s\le t<1650s$ & $1650s\le t<1925s$ & \textbf{Total} \\
\hline

\textbf{Percentages} 
  & 59\% & 54\% & 59\% & \textbf{52\%} \\
\hline

\multicolumn{5}{|l|}{
		\textbf{File:} OxIOD/running/data1/imu1\quad
	    \textbf{Raw data:} InvenSense ICM-20600\quad
		\textbf{Ground truth (same as{~\cite{russo2021danae++}}):} Vicon\quad
    } \\
\hline
\textbf{Parameters}
  & \multicolumn{4}{c|}{$Q_k=10^{-2}I_6$, $U_k=10^6\cdot I_6$, $\gamma = 1.0$} \\
\hline

\textbf{Intervals} 
  & $0\le t<40s$ & $40\le t<80s$ & $80s\le t<120s$ & $120s\le t<160s$ \\
\hline

\textbf{Percentages} 
  & 59\% & 21\% & 85\% & 71\% \\
\hline

\textbf{Intervals} 
  & $160s\le t<200s$ & $200s\le t<240s$ & $240s\le t<280s$ & \textbf{Total} \\
\hline

\textbf{Percentages} 
  & 52\% & 87\% & 83\% & \textbf{63\%} \\
\hline

\multicolumn{5}{|l|}{
		\textbf{File:} INSANE/mars\_1\quad
	    \textbf{Raw data:} Pixhawk IMU+Magnetometer\quad
		\textbf{Ground truth (same as{~\cite{brommer_insane_2022}}):} Pixhawk IMU+Magnetometer \quad
    } \\
\hline
\textbf{Parameters}
  & \multicolumn{4}{c|}{$Q_k=10^{-2}I_6$, $U_k=10^4\cdot I_6$, $\gamma = 1.0$} \\
\hline

\textbf{Intervals} 
  & $0\le t<15s$ & $15\le t<30s$ & $30\le t<45s$ & $45s\le t<60s$ \\
\hline

\textbf{Percentages} 
  & 41\% & 34\% & 61\% & 41\% \\
\hline

\textbf{Intervals} 
  & $60\le t<75s$ & $75\le t<90s$ & $90s\le t<105s$ & \textbf{Total} \\
\hline

\textbf{Percentages} 
  & 36\% & 29\% & 84\% & \textbf{43\%} \\
\hline
\end{tabular}
}
}
\smallskip
\caption{Positive score of Rev-MEKF for different datasets at different time intervals. The Ground truth used is extracted from the articles associated to the datasets. }
\label{tab:compare_datasets}
\end{table}
\end{center}

We deduce from Table~\ref{tab:compare_datasets} that the heuristic switches between Rev-MEKF and MEKF in specific scenarios.

First, for movements close to static movements, it correspond to time intervals of datasets with low percentages. It does not mean that Rev-MEKF does not correct on MEKF. For example, on industrial robot dataset, between $9$ and $12$ seconds, the correction are given in Figure~\ref{fig:diff_mekf_revmekf_instrual_robot}: although they are rare, their quality is better, as it is discussed in Section~\ref{sec:qualityanalysis}.

Moreover, the highest percentages are achieved in the running file from OxIOD dataset, which is a scenario where the external acceleration is constant and, consequently the quaternion obtained from MEKF has a constant bias when compared to the Ground Truth.

\begin{figure}[h!]
\centering
\includegraphics[width=\columnwidth]{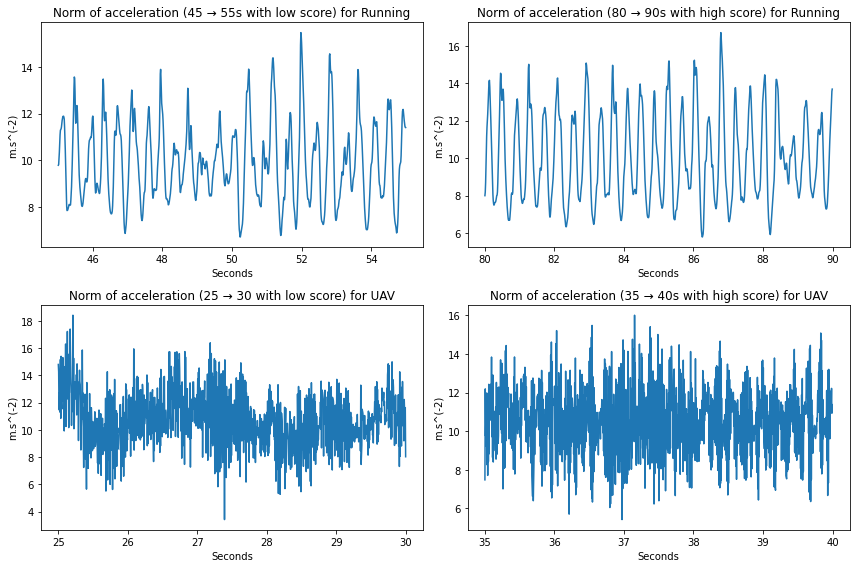}
	\caption{Norm of the acceleration on a low score packet on left, high score on the right (score refers to Table~\ref{tab:compare_datasets}). Top: OxIOD dataset, file: runnning/data1/imu1. Sensor: InvenSense ICM-20600. Bottom: INSANE dataset, file: mars\_1. Sensor:  Pixhawk IMU+Magnetometer.}
\label{fig:norm_acceleration_oxiod_uav}
\end{figure}

In Figure~\ref{fig:norm_acceleration_oxiod_uav}, we give, for OxIOD and INSANE dataset, a comparison of the acceleration on two intervals where the statistics are different, showing that the type of motion is different: acceleration is higher in average when the score of Rev-MEKF is higher. In particular, a measure of the variation of the acceleration and its amplitude is higher on intervals with better score.

We conclude that the score of the heuristic depend on the type of motion. The more there is amplitude in acceleration, the more Rev-MEKF brings corrections.
\subsection{Qualitative analyis}
\label{sec:qualityanalysis}
We select in this section a few moments coming from the cited datasets for which we have interesting examples since they allow one to test the behavior of a filter on simple trajectories, which simplifies our analysis. In particular, we give examples from industrial robot dataset~\cite{LowcomplexityMARG} (pure rotations) and Caves sonar dataset~\cite{10.1177/0278364917732838} (planar trajectory) to show that, with the metric proposed in Definition~\ref{def:metric}, one can show that the heuristic used on Rev-MEKF has a physicial meaning.

The industrial robot dataset~\cite{LowcomplexityMARG} allows one to compare reference sensors to a sensor providing raw data. In the example used in this section, we compare the quaternion obtained from a XSENS MTi-30 sensor at 100Hz, used as a ground truth in ~\cite{LowcomplexityMARG}, to the data obtained with an MPU9150, providing gyroscope, accelerometer and magnetometer data at 100Hz, but with a significantly lower precision.

%\begin{figure}[h!]
%\centering
%\includegraphics[width=0.75\columnwidth]{figures/gyro_industrial_robot.png}
%\caption{Gyroscope data of the industrial robot dataset, file V1000/Sequencia1.}
%\label{fig:gyro_industrial_robot}
%\end{figure}

\begin{figure}
\centering
\includegraphics[width=0.9\columnwidth]{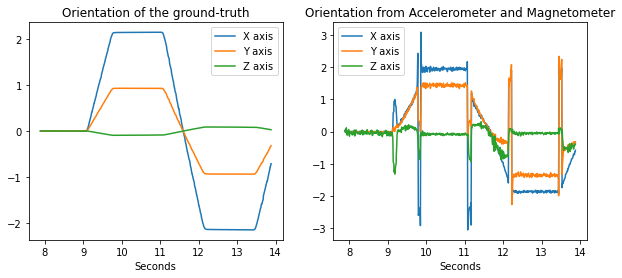}
\caption{Orientation of the ground truth on the left and orientation deduced from accelerometer and magnetometer on the right, in axis-angle representation. Industrial robot dataset, file V1000/Sequencia1.}
\label{fig:compare_quat_industrial_robot}
\end{figure}

In Figure~\ref{fig:compare_quat_industrial_robot}, we compare the axis angle representation of the orientation obtained via the Ground Truth during the rotation around $X$ axis and the orientation obtained from the accelerometer and magnetometer data.
The main advantage of this dataset is that it allows to test harsh dynamic environment. In Figure~\ref{fig:compare_quat_industrial_robot}, one observes what kind of motion we are considering: pure rotation around the $3$ axis $Y$, $Z$ and $X$.

%q0,q1,r0,r1 = 10**(-2), 10**(-2), 10**(2), 10**(2)

\begin{figure}[h!]
\centering
\includegraphics[width=0.85\columnwidth]{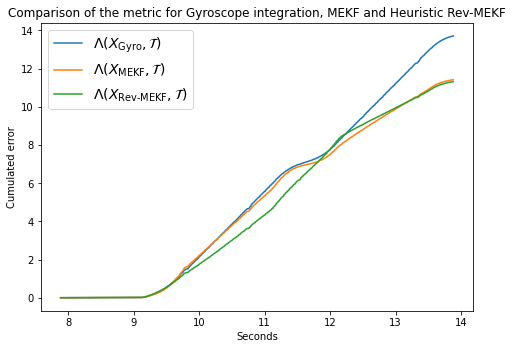}
\caption{Comparison of the metric for Gyroscope integration, MEKF and Heuristic Rev-MEKF on data provided by a MPU9150 sensor. Ground truth is the quaternion returned by an XSENS MTi30 sensor.  Industrial robot dataset, file V1000/Sequencia1.}
\label{fig:compare_variants_industrial_robot}
\end{figure}

We have been able, during the rotation around $X$ axis to compare the orientation obtained through the integration of the gyroscope, with the MEKF variant and the Heuristic Rev-MEKF.
The matrices used in prediction and update step are:  $Q_k = 10^{-2}\cdot I_{6\times 6}$ and $U_k = 10^{2}\cdot I_{6\times 6}$. The parameter $\gamma$ has been set to $1$ in Rev-MEKF.
Figure~\ref{fig:compare_variants_industrial_robot} compares the metric of these $3$ approaches. Gyroscope integration drifts over time, and Rev-MEKF improves on MEKF at the beginning.

\begin{figure}[h!]
\centering
\includegraphics[width=\columnwidth]{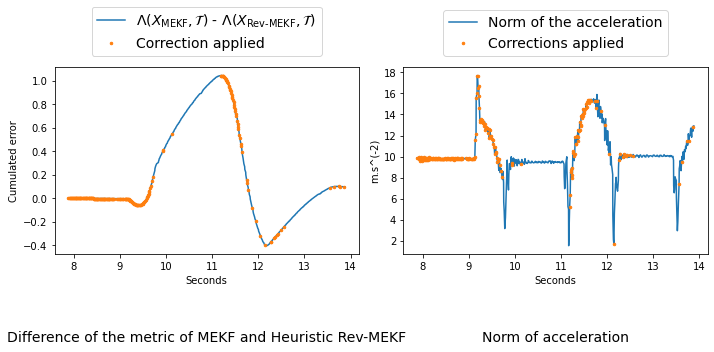}
\caption{Difference of metric $\Lambda(X_{\text{MEKF}},\mathcal{T})-\Lambda(X_{\text{Rev-MEKF}},\mathcal{T})$ on data from an MPU9150 sensor (ground truth from an XSENS MTi30 sensor) on the left, and norm of the acceleration on the right. Industrial robot dataset, file V1000/Sequencia1.}
\label{fig:diff_mekf_revmekf_instrual_robot}
\end{figure}

In Figure~\ref{fig:diff_mekf_revmekf_instrual_robot}, we compute the difference $\Lambda(X_{\text{MEKF}},\mathcal{T})-\Lambda(X_{\text{Rev-MEKF}},\mathcal{T})$.
The most interesting aspect of this figure is its relation to the accelerometer norm.
The metric of Definition~\ref{def:metric} allows us to identify moments when the Heuristic Rev-MEKF outperforms the standard MEKF. As expected, these correspond to periods of high external acceleration, visible on the right side of Figure~\ref{fig:diff_mekf_revmekf_instrual_robot}. These improvements appear at the beginning of the sequence (between $9$ and $10$ seconds) and at the end (between $12$ and $13$ seconds). In contrast, the correction applied between $11$ and $12$ seconds does not lead to improved performance.

We conclude that the heuristic is mainly activated during dynamic motion phases. Moreover, we have explicit examples of such dynamic motion where the metric of Rev-MEKF improves. In Figure~\ref{fig:diff_mekf_revmekf_instrual_robot}, the series of points between $9$ and $10$ seconds and between $11$ and $12$ seconds can serve as references to understand why, in similar situations, the metric is not improved. A next step in algorithmic development would be to parameterize the heuristic so that it correctly captures the dynamic motion occurring between $12$ and $13$ seconds.

By contrast, the underwater caves sonar dataset is useful because the motion is almost planar, which simplifies the comparison of different filter variants.

\begin{figure}
\centering
\includegraphics[width=\columnwidth]{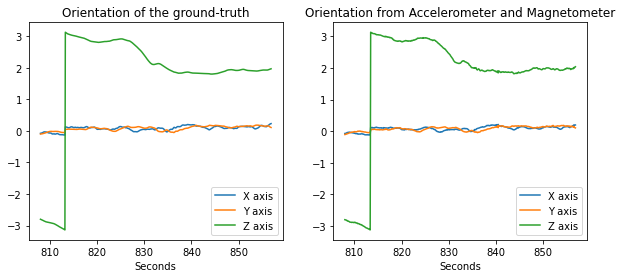}
\caption{Orientation of the ground truth (left) and orientation estimated from accelerometer and magnetometer (right), in axis--angle representation. Caves sonar dataset.}
\label{fig:compare_quat_cave_dataset}
\end{figure}

The matrices used in the prediction and update steps are $Q_k = 10^{-2} I_{6\times 6}$ and $U_k = I_{6\times 6}$. The parameter $\gamma$ in Rev-MEKF is set to 1.
Figure~\ref{fig:compare_quat_cave_dataset} shows the ground-truth orientation and the orientation estimated from accelerometer and magnetometer data. The two representations are similar and can be meaningfully compared. They also confirm that the motion is essentially planar, with variations occurring mostly in the yaw angle.

\begin{figure}[h!]
\centering
\includegraphics[width=\columnwidth]{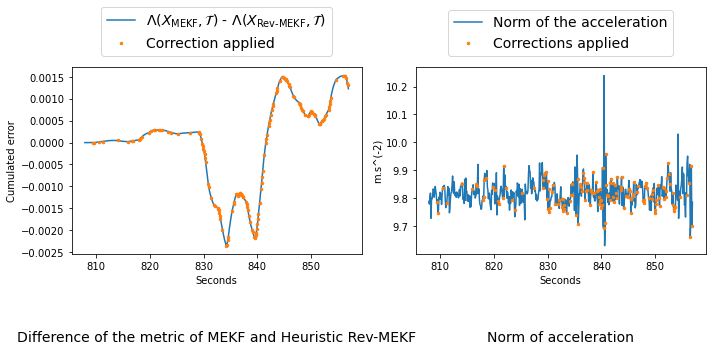}
\caption{Difference of metric $\Lambda(X_{\text{MEKF}},\mathcal{T})-\Lambda(X_{\text{Rev-MEKF}},\mathcal{T})$ on the left, and norm of the acceleration on the right. Caves sonar dataset.}
\label{fig:diff_mekf_revmekf_cave_dataset}
\end{figure}

In Figure~\ref{fig:diff_mekf_revmekf_cave_dataset}, we plot the difference $\Lambda(X_{\text{MEKF}},\mathcal{T}) - \Lambda(X_{\text{Rev-MEKF}},\mathcal{T})$.
As before, we focus on periods where this difference increases, as they can be related to changes in acceleration.
The metric difference shows that Rev-MEKF improves on MEKF between 835 and 845 seconds. Corrections occur more frequently in this dataset due to the slower motion. Nevertheless, some corrections clearly correspond to acceleration peaks, indicating that Rev-MEKF's improvements have a physical interpretation consistent with the behavior of the metric.

\section{Discussions}
\label{sec:discussions}

The following conclusions apply only to systems whose state is constrained to a plane.

\subsection{Testability of MEKF}

As discussed in~\cite{covanov2025reversiblekalmanfilterstate}, a primary use of Rev-MEKF is to assess the numerical precision of MEKF under extreme conditions. With arbitrary-precision arithmetic, one can observe a linear growth of numerical error over time, which allows the evaluation of different MEKF implementations.

This approach enables comprehensive testing for potential issues such as incorrect initialization, errors in left or right multiplication, or improper transition matrices. It also facilitates adapting the MEKF to new scenarios, including alternative residuals or sensor configurations.

\subsection{Adaptation to Real Data}

The metric defined in Definition~\ref{def:metric} makes it possible to identify moments when Rev-MEKF should be applied. This capability motivates the development of classifiers to detect when the heuristic Rev-MEKF is beneficial, providing a physical interpretation of the heuristic’s score. Such an approach can be extended to larger datasets, and the results could guide algorithms that more carefully select when to switch between Rev-MEKF and MEKF.

Moreover, the heuristic we studied can be interpreted as a generalization of a classification procedure based on the distance $h_k$. We believe that the distance in Definition~\ref{def:distance} offers a natural foundation for introducing classification within fusion filters. Since the heuristic is essentially a cost function based on the zeros of $h_k$, one could generalize this approach to alternative cost functions.

\subsection{Applications}

Incorporating a normal vector into the Kalman filter enables a broad range of applications. A primary example is vehicle localization using odometry and IMU sensors. This strategy is not limited to planar motion, since the normal vector may evolve over time.

One of the most promising applications lies in satellite or space trajectory estimation, where gyroscopes and accelerometers typically exhibit low noise and trajectories can be modeled on manifolds.

Autonomous vehicle navigation and wearable applications are also feasible, including human motion analysis during running, walking, or cycling, where motion can often be approximated as planar.

\bibliographystyle{IEEEtran}
\bibliography{bibtex/bib/bare_jrnl}
\end{document}